\documentclass{aa}
\usepackage{graphicx}
\usepackage{natbib}
\newcommand{\ROSAT}{{\it ROSAT}}
\begin{document}
   \title{Missing baryons and the soft X-ray background}

   \author{A. M. So\l tan \inst{1}, M. J. Freyberg \inst{2}
          \and
           G. Hasinger \inst{2}
          }

   \offprints{A. M. So\l tan}

   \institute{$^1$Nicolaus Copernicus Astronomical Center, Bartycka 18,
               00-716 Warsaw, Poland\\
              \email{soltan@camk.edu.pl} (AMS)\\
              $^2$ MPI f\"ur extraterrestrische Physik, Giessenbachstra{\ss}e,
               85748 Garching, Germany\\
              \email{mjf@mpe.mpg.de} (MJF)\\
              \email{ghasinger@mpe.mpg.de} (GH)
             }

   \date{Received July 16; accepted September 3, 2002}

  \abstract{
The X-ray background intensity around Lick count galaxies and rich clusters of galaxies
is investigated in three \ROSAT\ energy bands. It is found that the X-ray enhancements
surrounding concentrations of galaxies exhibit significantly softer spectrum than
the  standard cluster emission and the average extragalactic background.
The diffuse soft emission accompanying the galaxies is consistent with the thermal
emission of the hot gas postulated first by the Cen \& Ostriker hydrodynamic simulations.
Our estimates of the gas temperature - although subject to large uncertainties - averaged
over several Mpc scales are below $1$\,keV, which is substantially below the temperature
of the intra-cluster gas, but consistent with temperatures predicted for the local
intergalactic medium. It is pointed out that the planned {\it ROSITA} mission would
be essential for our understanding of the diffuse thermal component of the background.
   }
   \authorrunning{A. M. So\l tan et al.\ }

\maketitle

\section{Introduction}

The X-ray background (XRB) is mostly generated by discrete extragalactic sources (e.g.
\cite{lehmann01}, and references therein). Among those sources, various classes
of AGNs constitute a dominating part. Probably 5 to 10\,\% of the soft XRB is
produced by hot gas in clusters of galaxies. Around and below $1\,{\rm keV}$ hot plasma
in the Galaxy also contributes to the total background flux (\citealt{hasinger92}).
Apart from the source contribution, truly diffuse emission of extragalactic origin is
also expected. Using hydrodynamic simulations \cite{cen99} investigated
evolution of the primordial gas density and temperature. Baryons not condensed in
stars and interstellar medium within galaxies, occupy  the intergalactic space
and are spread over a wide range of temperatures and densities. This question is
discussed in detail by \cite{dave}, \cite{bryan} and \cite{croft}. According to all
the simulations, the hottest phase
is located in the high mass concentrations of clusters of galaxies and it is responsible
for the cluster X-ray emission. Relatively cold phase with temperature below $10^5$\,K
reveals its presence by the Lyman alpha forest.
The fraction which still escapes detection, described as {\it Warm-Hot}
Intergalactic Medium (WHIM), comprises of $30$ -- $40$\,\% of all baryons in the 
present-day universe (\citealt{dave}). Its temperatures are between $10^5$ and $10^7$\,K
and the hottest and high density fragments surround mass concentrations as groups and
clusters of galaxies. Thus, the thermal emission by WHIM should be easiest to detect
in the soft X-rays in the vicinity of rich clusters and high galaxy density areas.
However, as pointed by \cite{bryan}, compact sources
produce most of the background and observations put tight constraints on the level
of truly diffuse XRB component.

Detailed analysis of the diffuse emission produced by the intergalactic medium has been
presented by \cite{bryan} and \cite{croft}. Using hydrodynamic simulations they generated 
maps of the
X-ray sky which included contribution from the WHIM component. These maps have been used
to calculate the model autocorrelation function of the XRB at small angular scales (below
$\sim 10^\prime$) as well as the cross-correlation with the galaxy distribution.

In the present paper the distribution of the XRB in the vicinity of galaxy concentrations is
carefully investigated from the observational point of view. Lick counts (\citealt{shane},
hereafter SW\footnote{The Lick galaxy counts in $10^\prime$ pixels have been kindly
provided to us in the electronic form by Dr. M. Kurtz.}) and
Abell clusters of galaxies (\citealt{abell58}, \citealt{abell89}) have been used to select
areas of the high mass concentrations. Since the predicted WHIM temperatures are
substantially lower than those of the intra-cluster gas, and the WHIM emission is softer
that the average extragalactic XRB, the objective of our analysis is to look for
the systematic variations of the XRB spectral slope as a function of the distance from
peaks of the galaxy distribution. The expected WHIM contribution to the XRB
is small in comparison to the total background flux. To increase signal-to-noise ratio
we have measured the excess XRB flux around galaxy concentration using the
cross-correlation technique.  The X-ray data contained in the \ROSAT\ All-Sky Survey (RASS)
have been used.  For a comprehensive description of the RASS see \cite{snowden90},
\cite{voges} and \cite{snowden95}. 
Due to limited angular resolution of the RASS we have concentrated on
the larger scales than those investigated by \cite{croft}. In the next section the
basic concept of our investigation is presented and in Sect.~\ref{results} results of
calculations are given. We end our investigation with the short conclusions in the
Sect.~\ref{conclusions}.

\section{The Method \label{method}}

It is well established that the XRB exhibits fluctuations over a wide range of angular scales
(e.g. \cite{sliwa} and references therein). It is expected that the XRB variations
result from the nonuniform distribution of sources which generate the XRB (e.g.
\cite{soltan99}). One should note, however, that intrinsically large, extended sources also
produce fluctuations at the corresponding angular scales. Such fluctuations associated with
rich clusters of galaxies have been detected by \cite{soltan96}. They calculated
the cross-correlation function (CCF) between the distribution of the Abell clusters and
the RASS maps. The region apparently free from the strong galactic emission has been used.
The positive amplitude of the CCF extended up to $\sim 7^\circ$ and $\sim 3^\circ$ for
nearby and more distant clusters, respectively. The CCF signal indicated that clusters
are surrounded by halos of the enhanced X-ray emission. They estimated the average size
of the halo at approximately $20$\,Mpc. The amplitude of the X-ray enhancements
above the average XRB level is extremely low.
It amounts to about $1$\,\% of the total XRB surface brightness. Such weak
signal could not be detected using an individual cluster observation but it was
identified in the statistical sense in the cluster population
using the cross-correlation analysis.

Two explanations have been proposed in the original paper: discrete and diffuse. In the discrete
model the excess emission is produced by a large number of sources. The X-ray enhancements
result from the fact that the spatial density of galaxies around clusters is larger.
This increased density extends up to at least $10$\,Mpc from the clusters
(\citealt{seldner}). Such concentrations of galaxies naturally generate excess X-ray
volume emissivity, associated with the X-ray emission of normal galaxies and weak AGN
sources populating those galaxies. In the diffuse model it was {\it ad hoc} assumed that
the X-ray emission was produced in the extended hot clouds of gas surrounding the clusters. 

These two models demonstrate significantly different spectral properties.
As shown by \cite{cen99} and \cite{dave} the WHIM which would be a source
of the X-ray enhancements contributes mostly at and below $1\,{\rm keV}$, while 
the normal galaxy and AGN emission is substantially harder and is expected to have
spectral shape not distinctly different from the integrated XRB flux. The hot
intra-cluster gas with ${\rm k}T$ typically between $2$ and $10\,{\rm keV}$
also produces a 
harder spectrum than the WHIM. 

The CCF of the RASS and the SW counts has been determined in \cite{soltan97}. 
The clear correlation signal was interpreted as a result of the clustering
of galaxies, but some
indication of the extended emission associated with galaxies has been reported there.

The excess emission $\delta\rho(\theta)$ at angular distance $\theta$ from
the randomly chosen galaxy or cluster is related to the amplitude of the corresponding
CCF, $w(\theta)$, in the following way:
\begin{equation}
\delta\rho(\theta) = \langle\rho\rangle\, w(\theta),
\label{delrhoxi}
\end{equation}
where $\langle\rho\rangle$ is the average XRB signal. The details of the CCF calculations
are given in \cite{soltan96}. Here we recall the main points. The final RASS data are binned into 
$12^\prime\times12^\prime$ pixels. We have binned the galaxy/cluster samples in the same
way and the CCF was estimated using the formula:
\begin{equation}
w(\theta) = {{1\over m(\theta)}\,\sum_{\rm ij} \rho_{\rm i} n_{\rm j} \over 
            \overline{\rho}\, \overline{n}} -1,
\end{equation}
where $\rho_{\rm i}$ and $n_{\rm j}$ denote the X-ray count rate in the $i$-th pixel and
number of galaxies/clusters in the $j$-th pixel, respectively; $m(\theta)$ is the number
of pixel pairs with the separation equal to $\theta$ and the sum extends over all such
pairs; $ \overline{\rho}$ and $\overline{n}$ are the mean values of the XRB and galaxy/cluster
distributions.

The objective of the present calculations is to
determine the X-ray excess flux, $\delta\rho(\theta)$, as a function of the photon energy.
Unfortunately, the energy range of the \ROSAT\ PSPC useful in the present investigation
covers a rather narrow energy range, roughly between $0.7$ and $2.0\,{\rm keV}$
\footnote{The soft band below $0.28\,{\rm keV}$ is highly contaminated by local galactic
emission (\citealt{snowden95}).}.
In the calculations we have used three standard \ROSAT\ energy bands: R5, R6 and R7
with peak response at $0.8$, $1.1$ and $1.5\,{\rm keV}$, respectively (see \citealt{snowden94}
for details).
Since the PSPC energy resolution is relatively poor, those bands overlap what additionally
reduces the sensitivity of our analysis.

\section{Results \label{results}}

In the present calculations we have used the data at high galactic
latitude in the northern hemisphere ($b > 40^\circ$), where the extragalactic signal
seems to be least contaminated by the galactic emission: $70^\circ < l < 250^\circ$
(\citealt{soltan96}). Although in this region some local contribution is still present,
the CCFs between the RASS maps and optical samples are determined with relatively high
accuracy.

Since the present calculations are based on the angular cross-correlations, the effect
depends on the galaxy distances. The redshift distribution of SW galaxies is relatively wide
with a median value of $0.068$ and $80$\,\% of galaxies have redshifts between
$0.045$ and $0.17$. These estimates are based on extrapolations of the redshift distribution
of the early SDSS data (\citealt{dodelson}).

The Abell clusters (\citealt{abell58}, \citealt{abell89}) have been divided into two
distance groups. Sample I contains clusters with Distance Class $1$ through $4$ and all
Richness Classes, and sample II contains DC = $5$ and RC $\ge 1$. In the investigated
area samples I and II consist of 57 and 210 clusters, respectively. In our analysis
both samples give consistent results for all three energy bands (see below).
However, due to the small number of clusters in the sample I, uncertainties of our
measurements for that sample are too large to draw restrictive conclusions on the spectral
properties of the X-ray halos. For this reason, the subsequent analysis is limited
to the sample II. Using the \cite{abell89} data it is estimated that $80$\,\% of DC $=5$
clusters have redshifts between $0.07$ and $0.17$.

The SW counts and Abell clusters cover comparable redshift ranges and the distributions
of objects in both data sets are strongly correlated. To make the cross-correlation
analysis of the SW and RASS data independent of the Abell clusters, all the calculations
involving the SW galaxies were performed using a mask which removed the galaxies and
X-ray flux related to clusters 
\footnote{Radius of the masked out area depends on the cluster distance. For the nearest clusters
(DC $=1$) pixels within $48^\prime$ from the cluster position are removed, while for DC $\ge 5$
only pixels centered on the cluster and 8 neighbouring pixels are not used.}.

The results of our calculations are shown in Figs. 1 through 4 in the form of 
X-ray `colour-colour'
diagrams using different combinations of bands R5, R6 and R7. To see effects of gas
temperature variations with distance from galaxy concentrations, the excess fluxes
have been calculated using the corresponding CCF amplitudes at different separations.
Data representing clusters are shown with squares while the data referring to SW galaxies --
with pentagons.

\begin{figure}
\centering
\includegraphics[width=0.8\linewidth]{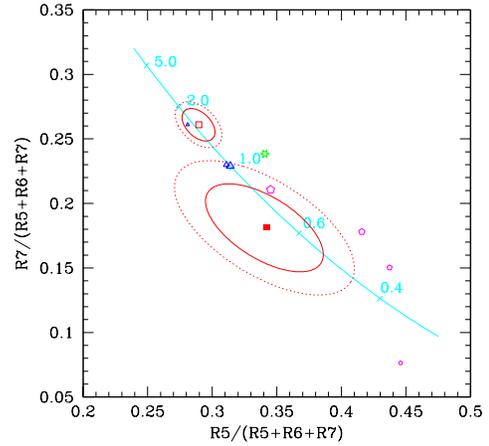}
\caption{The X-ray `colour-colour' diagram of the RASS data. Count rates in the
energy bands R5 and R7 normalized to the summed rates in the bands R5, R6 and R7
are used. Open square indicates colours of the excess emission produced by clusters
of galaxies; the full square - colours of the excess `halo' emission integrated 
between $0\fdg3$ and $2\fdg1$ from the clusters.
The ellipses show $68$\% and $90$\% confidence limits.
Pentagons indicate colours of the excess emission correlated with the SW galaxies at 4
separation bins: $< 0\fdg3$, $0\fdg3 - 0\fdg7$, $0\fdg7 - 1\fdg5$, and $1\fdg5 - 3\fdg1$,
where the smaller symbol corresponds to the larger separation;
triangles show colours of AGNs in three redshift bins:  $0.1 - 0.4$ (large symbol),
$0.4 - 1.0$ (medium), $1.0 - 2.0$ (small),  star - average
RASS colours calculated from the count rates integrated over all the investigated area.
Solid curve represents thermal Bremsstrahlung;
labels indicate the temperature in keV. See text for details.  \label{r5_7_th}}
\end{figure}

In Fig.~\ref{r5_7_th} the distribution of excess count rates in the R5 and R7 bands
normalized to the cumulative count rate excess in all three bands is shown. 
The empty square representing the central cluster regions shows the
emission determined from pixels containing the cluster center;
the full symbol -- represents the `halo' region which is defined
as an annulus with the inner and outer radius of $0\fdg3$ and $2\fdg1$.
The solid and dotted ellipses denote $68$\% and $90$\% confidence areas. 
The SW data are shown for 4 separation bins: the largest symbol denotes signal for
separations smaller than $0\fdg3$, and the next three pentagons of decreasing size
represent three separation bins of
$0\fdg3 - 0\fdg7$, $0\fdg7 - 1\fdg5$, and $1\fdg5 - 3\fdg1$, respectively.

To assess effectiveness and to inquire into potential systematic effects of our method,
we have calculated also excess emission associated with the population of AGNs listed in the
Veron catalogue (\citealt{veron}).

\begin{figure}
\centering
\includegraphics[width=0.8\linewidth]{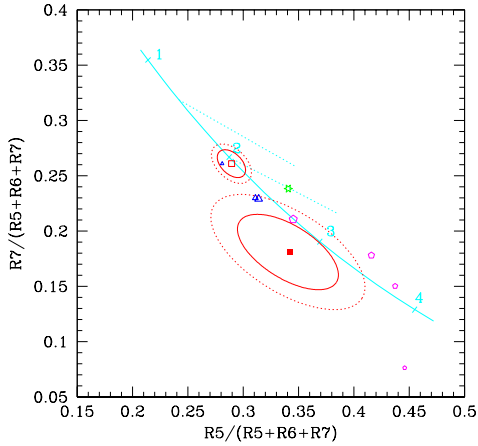}
\caption{All data points - same as Fig.~\ref{r5_7_th}.
Solid curve represents colours of the pure power law models; labels denote the photon
indices. Two dashed lines indicate loci of the two component models. The upper line: power law
with index $\Gamma = 1.4$ with the addition (up to $10$\,\% in the R6 band) of the thermal
Bremsstrahlung with ${\rm k}T = 0.1\,{\rm keV}$; the lower line - the same for
$\Gamma = 2$. \label{r5_7_pl}}
\end{figure}

Thermal Bremsstrahlung spectra are given by the solid curve; labels
denote temperatures in keV. In Fig.~\ref{r5_7_pl} the same data points
are superimposed
on the solid curve indicating the location of pure power law spectra; 
marks with labels
give the photon indices. Potential effects of the Local Bubble and our own Galaxy
halo emission on the integrated background flux are shown with two dashed lines.
The lower line represents the sum of the power law spectrum
with the photon index $\Gamma = 2$ and the thermal Bremsstrahlung with 
${\rm k}T = 0.1\,{\rm keV}$.
The contribution of the thermal component increases along the line from $0$
at the left-hand end of the line to $0.1$ of the counts in the R6 band at the right-hand
end. The upper line represents spectra with $\Gamma = 1.4$ and the thermal
component varying in the same range.

Uncertainty ellipses for the cluster sample have been delineated using the distributions
of 1000 simulations.  In each simulation, the count rates in randomly chosen RASS pixels
were used to perform analogous computations as for the real data. In each simulated run
the number of drawn pixels was equal to the number of clusters in the sample. Because the
halo signal is substantially weaker than the `core' signal, our estimates of the halo
temperatures are much less accurate.

The error ellipses for the core and halo emission containing more than $\sim 94$\,\% of
the simulation runs begin to overlap. This puts the lower limit for the significance
of the temperature difference for these two cluster-centered regions. The statement
is strengthened by the fact that the effect is present for any combination of the X-ray
colours (see the remaining figures below).

In the Figs.~\ref{r5_6_th} and \ref{r6_7_th} the distribution of all the categories
of objects is shown using different combinations of the \ROSAT\ bands.
Although both SW and cluster points exhibit large uncertainties and scatter,
all the data display a common feature. Namely, the emission parametrized
by the temperature (or the spectral slope) correlated with the cluster and SW galaxies
distribution becomes systematically softer as the distance from the galaxy
concentration increases.

The results for the cluster sample can be summarized as follows: the cluster flux is 
consistent with the thermal emission with $\rm{kT} \approx 1.5 - 2\,{\rm keV}$ (see below)
and the halo flux - with ${\rm k}T \approx 0.5 - 0.8$\,keV. Although 
halo uncertainties are large and this temperature difference is barely significant,
the general trend in all the `colour-colour' diagrams is highly suggestive.
Similar temperature differences between the core and halo excesses derived for
all combinations of the energy bands demonstrate that the distribution of count rates in
each band contributes to the resultant temperature effect around the cluster population.
The excess associated with the cluster halos appears to be consistent with thermal emission
generated by the WHIM.

The excess emission correlated with the SW galaxies is still softer than that around clusters
indicating lower gas temperature in qualitative agreement with simulations.
Quantitative assessments of this effect are hindered by systematic effects (see below)
as well as large statistical errors. 
Although in the case of the SW points alone systematic colour variations
with the increasing separation are of low statistical significance,
combined analysis of the cluster and SW samples shows that the effect is real.
The error bars (not shown in figures for clarity) for three colours of
the `zero lag' SW data representing separations below  $0\fdg3$
are in the range of $0.018 - 0.022$. Thus, the SW signal is distinctly softer
than the cluster `core' emission. At larger separation the CCF amplitude
suffers from substantially larger uncertainties and consequently
colours of the `$1\fdg5 - 3\fdg1$' bin differ from the `$<0\fdg3$'  bin
by just over $1\,\sigma$. One should note however,
that the resultant CCF amplitude is a superposition of signals generated
by the galaxy clustering and the extended emission surrounding galaxies. This conclusion
is supported by our earlier analysis of the SW -- RASS correlation (see \citealt{soltan97}).

Our estimates of the CCF between the RASS maps and the SW galaxies cannot be directly
compared with the CCFs generated by \cite{croft} for two reasons. Galaxies simulated
by Croft et al.\  are more distant than the SW galaxies and, what is more important, those
simulations are not devised to reproduce the discrete emission by galaxies,
while in our measurements the CCF amplitude represents the total correlation signal.
In the present investigation the IGM emission is revealed because it is substantially
softer than the galactic emission.

\begin{figure}
\centering
\includegraphics[width=0.8\linewidth]{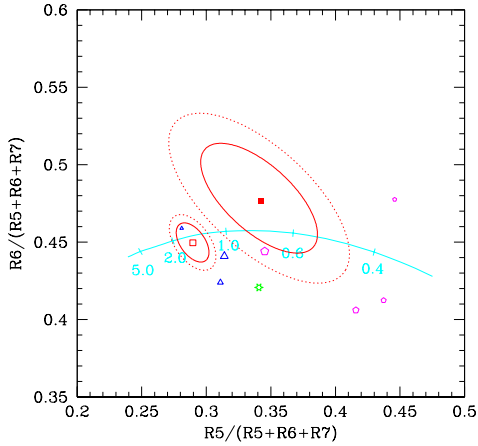}
\caption{Same as Fig.~\ref{r5_7_th} for the count rates in the bands R5 and R6.\label{r5_6_th}}
\end{figure}

\begin{figure}
\centering
\includegraphics[width=0.8\linewidth]{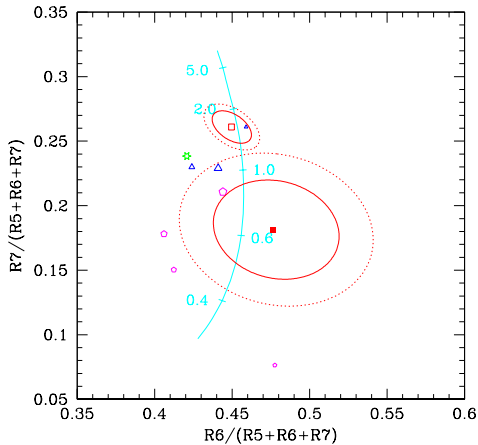}
\caption{Same as Fig.~\ref{r5_7_th} for the count rates in the bands R6 and R7.\label{r6_7_th}}
\end{figure}

It is disturbing that our estimate of the average cluster temperature falls below
$2\,{\rm keV}$. This value is significantly lower than the median temperature
of $\approx\!5\,{\rm keV}$ observed in rich clusters (e.g. \citealt{david},
\citealt{horner}). This discrepancy could be removed by a measured adjustment
of the instrument calibration. However, modification of the effective area in bands
R5, R6 and R7 which are required to remove the cluster temperature inconsistency
are of the order of $10-15$\,\% what is larger than residual uncertainties of the
X-ray telescope/PSPC parameters (see below). In the recent study of the cluster X-ray
emission distribution \cite{bonamente} report detection of soft excesses in a large
fraction of sources. The authors claim that in the soft \ROSAT\ band of
$\sim 0.25$\,keV detected fluxes are on the average $9$\,\% larger than
it is expected from extrapolation based on measurements at higher energies
(above $\sim 2$\,keV). 

To test if this phenomenon could explain our low temperature cluster
measurement, we have calculated `colours' of the X-ray emission (defined by the R5, R6
and R7 bands) for the two temperature thermal Bremsstrahlung. Small amount of the low
temperature emission was added to the main `standard' component of $5$\,keV.
It was found that $\sim 0.7$\,keV emission normalized to $9$\,\% in the
low energy band used by \cite{bonamente} affects the R5, R6 and R7
colours in such a way that they mimic the thermal emission with the
temperature of $2 - 2.5$\,keV. Thus, this effect only partially
explains our cluster temperatures, but the discrepancy is substantially
reduced. 

A question of the `soft excess' is complicated further by systematic
uncertainties of the instrument calibration below the carbon edge (i.e. $0.28$\,keV).
The comparison between PSPC, Chandra and XMM soft sources shows that the PSPC
gives about $20-30$\,\% larger absolute fluxes at these energies. The agreement
above $0.5$\,keV is much better ($5-10$\,\%). Thus, the present investigation
is practically unaffected by uncertainties in the soft bands since the  R5 -- R7
bands have negligible responses below $0.5$\,keV (\citealt{snowden94}), but
the \cite{bonamente} result could be subject to calibration errors.

Systematic errors of the X-ray 'colours' in the RASS indicated by the cluster
'core' temperature are of the order of $0.05$. Although we are unable to point out
rigorously a single cause of this error, no potentially responsible effect could
alter the main conclusion of the present investigation, that the excess emission
correlated with the galaxy distribution softens with the increasing size of the
investigated area. Systematic errors in the 'colour - temperature' relationship
introduce uncertainty of the cluster 'halo' temperature at the level of
$0.1 - 0.2$\,keV, and in the case of large separations in the SW galaxies - of
about $0.1$\,keV. Consequently, all our temperature determinations should be
increased correspondingly. Thus, the cluster `core' temperature
disagreement and statistical noise precludes us from making a quantitative
determination of the IGM temperature.  But our investigation shows
convincingly that concentrations of galaxies are surrounded by emission of
significantly lower temperature than the average XRB and the cluster 'core' emission. 

One should note that the typical surface brightness of the IGM in the cluster halo
is roughly two orders of magnitude lower than the signal measured in pixels centered
on clusters. Thus, a halo contribution to the amplitude of the CCF at zero lag is
negligible. It implies that our estimate of the core temperature remains practically
unaffected by the halo component.

\section{Summary and conclusions \label{conclusions}}

A perceptible fraction of the XRB fluctuations is correlated with rich clusters
of galaxies and with the overall galaxy distribution. The cross-correlations
between the XRB and Abell clusters/SW galaxies extends up to several degrees.
The large angular size enhancements of the XRB which surround regions of higher
galaxy densities have distinctly softer spectrum than the genuine cluster emission. 
It is highly suggestive that these enhancements are generated by the WHIM.
The halo emission is consistent with the thermal Bremsstrahlung with ${\rm k}T$ below $1$\,keV.
Taking into account large uncertainties of the present estimates and a wide range
of plasma temperatures predicted by \cite{cen99} and \cite{croft}, our results are
in good agreement with theoretical models. Limited energy range used in this
investigation combined with the moderate energy resolution of the PSPC data prevent
us from more detailed conclusions.

This limitation applies also to our evaluations of the AGN spectral characteristics.
Although the average X-ray colours of a population of AGNs are adequately mimicked
by the thermal spectrum (Fig.~\ref{r5_7_th}), simple power law spectral models
also provide good fits (Fig.~\ref{r5_7_pl}). 
Uncertainties (not shown in figures) of the AGN colours are of the order of
$0.009$, $0.015$, $0.050$ for redshift bins $0.1 - 0.4$, $0.4 - 1.0$, $1.0 - 2.0$,
respectively. Thus, all three samples do not exhibit significant colour differences.
The majority of AGNs used in the present analysis are luminous X-ray
sources and their average spectral properties are not identical with the average spectrum
of weak sources which generate most of the XRB. Additionally, the integral soft XRB
is contaminated by thermal emission of our Galaxy (\citealt{hasinger92}). Importance
of this effect in our calculations is illustrated in Fig.~\ref{r5_7_pl} with dashed lines.

Most if not all shortcomings of the present investigation would be eliminated
with the {\it ROSITA} mission which is being proposed. The {\it ROSITA} telescope
mounted on board of the International Space Station would perform an all-sky survey
within the energy range of $0.5-10$\,keV and a sensitivity about 100 times
better than the RASS. These characteristics are particularly suitable for the
analysis of the low surface brightness features generated by the WHIM.
Such investigation  requires large, unlimited field of view. Also
wide energy range extending well above the WHIM domain would help to isolate
the WHIM thermal emission from the non-thermal components
of the X-ray emission correlated with the galaxy distribution. High sensitivity
would substantially increase efficiency of the correlation method by reducing
the photon noise statistics.

Finally, we would like to note that although the extended clouds of hot gas around
clusters in principle are sources of the Sunyaev-Zel'dovich (S-Z) effect (\cite{sunyaev}),
simple estimates by \cite{soltan96} and
\'Sliwa et al.\  (in preparation) indicate that for temperatures of the gas
lower than $1\,{\rm keV}$, the amplitude of the S-Z effect generated by halos is 
below the detection threshold of the {\it COBE} DMR measurements.

\vspace{2mm}
ACKNOWLEDGEMENTS. The \ROSAT\ project has been supported by the 
Bundesministerium f\"ur Bildung, Wissenschaft, Forschung und Technologie
(BMBF/DARA) and by the Max-Planck-Gesellschaft (MPG).
This work has been partially supported by the Polish KBN grant 5~P03D~022~20.

\end{document}